\documentclass[twocolumn,superscriptaddress,preprintnumbers,amsmath,amssymb]{revtex4}

\usepackage{graphicx}
\usepackage{dcolumn}
\usepackage{amsmath}
\usepackage[perpage,symbol*]{footmisc}

\makeatletter
\def\btt#1{\texttt{\@backslashchar#1}}%
\DeclareRobustCommand\bblash{\btt{\@backslashchar}}%
\makeatother

\begin{document}

\title{Can Natural Sunlight Induce Coherent Exciton Dynamics?}

\author{ Jan Ol\v{s}ina}
\email{olsina@karlov.mff.cuni.cz}
\affiliation{Department of Chemistry, Massachusetts Institute of Technology, Massachusetts, 02139 USA}
\affiliation{Faculty of Mathematics and Physics, Charles University in Prague, Ke Karlovu 5, CZ-121 16 Prague 2, Czech Republic}

\author{Arend G. Dijkstra}
\affiliation{Department of Chemistry, Massachusetts Institute of Technology, Massachusetts, 02139 USA}

\author{Chen Wang}
\affiliation{Department of Chemistry, Massachusetts Institute of Technology, Massachusetts, 02139 USA}
\affiliation{Singapore-MIT Alliance for Research and Technology, 1 CREATE Way, Singapore 138602, Singapore}

\author{Jianshu Cao}
\email{jianshu@mit.edu}
\affiliation{Department of Chemistry, Massachusetts Institute of Technology, Massachusetts, 02139 USA}

\date{\today}
\widetext

\begin{abstract}

Excitation of a model photosynthetic molecular aggregate by incoherent sunlight is systematically examined.
For a closed system, the excited state coherence induced by the sunlight oscillates with an average amplitude that is inversely proportional to the excitonic gap,
and reaches a stationary amplitude that depends on the temperature and coherence time of the radiation field.
For an open system, the light-induced dynamical coherence relaxes to a static coherence determined by the
non-canonical thermal distribution resulting from the entanglement with the phonon bath.
The decay of the excited state population to the  common ground state establishes a non-equilibrium steady-state flux driven by the sunlight,
and it defines a time window to observe the transition from dynamical to static coherence.
For the parameters relevant to photosynthetic systems, the exciton dynamics initiated by the sunlight exhibits a non-negligible amount of dynamical coherence (quantum beats) on
the sub-picosecond timescale; however,  this sub-picosecond time-scale is long enough for light-harvesting systems to establish static coherence, which plays a crucial role in
efficient energy transfer.   Further, a relationship  is established between the non-equilibrium steady-state induced by the sunlight and the coherent dynamics initiated from the
ground state by a laser $\delta$-pulse, thereby making a direct connection between incoherent sunlight excitation and ultrafast spectroscopy.
\end{abstract}


\maketitle \narrowtext

\section{Introduction}

Recent developments in 2D electronic spectroscopy (2DES)
of molecular aggregates have
demonstrated the presence of coherent dynamics in the electronic
degrees of freedom (DOF), known as electronic coherence, as opposed to
the expected incoherent hopping dynamics.
2DES was first used in the Fleming group to measure long-lived
oscillations in the Fenna-Matthews-Olson (FMO) photosynthetic pigment-protein
complex, which reveals  the presence of long-lived
electronic coherence \cite{Engel2007a}. This has led to a heated debate
about its importance to the efficiency of the excitonic energy transfer
in photosynthesis in general. Since
then, long-lived coherence has been reported in many other systems \cite{Engel2007a,Mercer2009a,Collini2010a} .
The term quantum coherence, generally used to describe the off-diagonal
elements of the density matrix in the exciton basis,
can be associated with several physical
phenomena. It is particularly important here to
distinguish between a \textit{static coherence} that remains constant
for long times and corresponds to stationary effects in equilibrium states or
non-equilibrium steady states  (e.g.,  localization and entanglement with the
bath), and a \textit{dynamical coherence}, which
is a transient effect associated with the time-evolution of the superposition of eigen states.
The latter is related to the recently discovered
quantum beats in 2DES measurements, which can be quantified as transient oscillations in
the coherence term.    The issue of sunlight-induced coherence
refers to the 'dynamic coherence' associated with the quantum beats;  whereas the entanglement with the phonon or photon bath leads to 'static coherence'
 associated with non-canonical thermal distributions.

One of the important questions in this discussion is whether electronic
coherent dynamics can be initiated by natural sunlight \cite{Jiang1991a,Mancal2010a}.
In 2DES or pump-probe experiments, molecules are excited by ultra-short
femtosecond laser pulses which lead to a pure quantum
state -- generally in the form of a superposition of several locally excited
states. On the other hand, natural thermal light is incoherent:  In
the semi-classical picture, it can be described as uncorrelated noise
that leads to an excitation of populations, but not necessarily coherence. In
the quantum picture, it was argued that the arrival of photons
does not have to be localized at a particular time nor have a well-defined
phase -- the thermal light behaves more like a dissipative quantum
bath at high temperature with a very short coherence time ($<10\;\mathrm{fs})$.
Although there have been many models of
the incoherent light excitation \cite{Jiang1991a,Mancal2010a},
the exact quantum description of the light, as well as a thorough analysis
of its properties over a broad range of parameters, is still not
available. This article aims to fill this gap. We use the Hierarchical
Equations of Motion (HEOM)  \cite{Tanimura1989a,Ishizaki2009a,Dijkstra2010,Moix2013}
as a non-perturbative description of both the phonon bath and photon
radiation.  We further introduce a decay channel, which
defines an observation window for the induced exciton dynamics.
The Drude-Lorentz spectral density assumed by the HEOM differs
from the black-body spectrum; however, in contrast to the classical
approach,  the HEOM correctly describes temperature effects and
all the relevant quantum effects and allows us to examine the
dependence of the induced dynamics on the coherence of the radiation field.

In the first part, we analyze the dynamics of a closed
system pumped by incoherent light. The amount of dynamical coherence (i.e, the amplitude of oscillations) generated under sunlight pumping
is constant, inversely proportional to the exciton energy gap,  but decreases if normalized by the linearly growing exciton populations.
We analyze the amount of  dynamic coherence for different coherence times of
the radiation and show that for very incoherent radiation, which is
the case relevant for the natural sunlight, the white-noise model (WNM)
provides a reliable description of the optical excitation. Open quantum systems
allow the additional dephasing mechanism, which depends on the detailed
properties of the system-bath coupling. Our analysis of the closed system aims to
demonstrate that the presence of  dynamical coherence is not excluded
by the fact that solar light is incoherent, provided the additional
dephasing of the electronic coherence from the phonon bath
is sufficiently slow, as observed in many biological systems.
Further analysis of the open system reveals
the phonon-induced static coherence, which appears as the long time limit
after the dynamic coherence is suppressed by the phonon noise.

In the second part, we include a decay channel which sets a time-scale for experimental observation
on the individual molecule level. Physically,  the decay channel can be interpreted as trapping at the reaction center,
fluorescence emission, or non-radiative decay~\cite{Cao2006,Shapiro2013}.  The density matrix formalism is a general tool to evaluate
experimental results in the framework of dissipative quantum dynamics.
However, it does not necessarily tell us about the state of individual molecules, which
is an issue closely related to the measurement problem in single molecule experiments.
The statistical interpretation of the density matrix formalism offers a mechanism
in which the pure quantum state of individual molecules can survive
for long times, and the observed loss of coherence is attributed
to ensemble averaging or time averaging, which results in cancelation of phase
coherence in states with different quantum phases.  This type of decoherence often applies to
dynamical coherence.    In contrast,  because of the coupling to phonons,
the coherence is lost even on the level of individual
molecules after the trace over the bath is performed.
This type of decoherence mainly applies to static coherence.
 We cannot differentiate these two types of coherence based on
the density matrix description but can distinct them on the single molecule level,
which will be further discussed in a future publication.
For example, in ensemble measurements, there is not a  particular
event to set time zero in our experiment,  and one thus has to be careful
about the initial condition~\cite{Cao2006}.   We try
to avoid these interpretational issues by introducing a decay channel,
which introduces a natural time-scale for energy transfer and establish
the non-equilibrium steady state.  Then, the contribution of light-induced dynamical coherence
to light-harvesting energy transfer is determined by the ratio of the dephasing time
and decay time, and may not play a dominant role in light-harvesting
systems, such as FMO or LH2.


\begin{figure}
\begin{center}
\centerline{\includegraphics[scale=0.6]{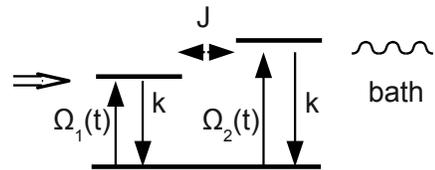}}
\caption{ Schematic V-type model with two excited states and
the common ground state, which are coupled with the radiation field of the sunlight,
with independent phonon baths,  and with a decay rate $k$ to the ground state.  The excited
states are coupled by a resonance constant $J$.} \label{fig:System-Schema}
\end{center}
\end{figure}

\section{Model and Methods\label{sec:Results-Discussion}}

\subsection{Molecular Dimer System}

The system of interest is a molecular dimer in contact with
 sunlight and two independent phonon baths,
which represent both vibrational DOF of the
molecules and the nuclear DOF of the surrounding environment (see Fig.~\ref{fig:System-Schema}).
This model captures the essential physics of delocalized exciton states
and coherent exciton dynamics relevant for light-harvesting systems.

The molecular dimer system has been extensively studied in quantum information, quantum optics,
and F\"{o}rster energy transfer theory~\cite{Moix2013}.
Each molecule, denoted by an index $i\in\{1,2\}$, is either in its electronic ground state
$|g_{i}\rangle$ or its electronic excited states $|e_{i}\rangle$ (i.e., excitons).
For convenience, we introduce the composite states $|\bar{g}\rangle=|g_{1}\rangle|g_{2}\rangle$,
$|\bar{e}_{1}\rangle=|e_{1}\rangle|g_{2}\rangle$, $|\bar{e}_{2}\rangle=|g_{1}\rangle|e_{2}\rangle$
and $|\bar{f}\rangle=|e_{1}\rangle|e_{2}\rangle$.   We remove the $|\bar{f}\rangle$-state
from our model since its influence is negligible due to its high energy.
Then, the molecular dimer can be described by the system Hamiltonian
\begin{eqnarray}
\hat{H}_S=
\begin{pmatrix}
\epsilon_1 & J & 0\\
J & \epsilon_2  & 0\\
0 & 0  & \epsilon_g\\
\end{pmatrix}
\end{eqnarray}
where the energies of states $|\bar{e}_{1}\rangle$, $|\bar{e}_{2}\rangle$,
$|\bar{g}\rangle$ are denoted $\epsilon_{1}$, $\epsilon_{2}$, $\epsilon_{g}$, respectively.
The symbol $J$ denotes the exciton resonance-coupling.
The two-level-system Hamiltonian in the excited state manifold can be diagonalized in the excitonic basis consisting of $|\bar{e}_+\rangle$ and $|\bar{e}_-\rangle$.
Throughout this paper,
we use the subscripts $\{1,2\}$ to denote the molecular (local) basis and subscripts $\{+,-\}$ to denote the excitonic basis, and
consider the off-diagonal matrix element $\rho_{+-}$  in the excitonic basis as a measure of coherence.
The typical parameters for light-harvesting systems are  $\epsilon_{1}=9900\;\mathrm{cm}^{-1}$,
$\epsilon_{2}=10000\;\mathrm{cm}^{-1}$, $\epsilon_g=0$,  and $J=50\;\mathrm{cm}^{-1}$.
For a closed system, the physical pictures for arbitrary $J$ are equivalent.
Hence, we can simply take $J=0$ for convenience, without losing any generality


The light-matter interaction between the molecular dimer and the radiation field
occurs via the dipole coupling, given by
\begin{eqnarray}~\label{HSR-article}
\hat{H}_{S-R}&=&-\hat{\mu} {\otimes} \hat{E}\\
&=&-\sum_{v=1,2}[\mu_{v}|\bar{e}_v\rangle\langle \bar{g} |+H.c.] {\otimes} \hat{E}\nonumber
\end{eqnarray}
where $\hat{\mu}$ is the total transition dipole operator of the dimer, and $\hat{E}$ is the radiation field, which are briefly discussed in the Appendix A.
In order to simplify the formulation and unify the formalism with the phonon baths, we rewrite the light-matter interaction term
as  $\hat{H}_{S-R}=\hat{K}_R{\otimes}\hat{V}_R$,  where $\hat{K}_{R}=\hat{\mu}/\mu$ and $\hat{V}_{R}=-\mu \hat{E}$.
The constant $\mu=\frac{1}{2}\sqrt{\mathrm{Tr}\,\hat{{\mu}}\cdot\hat{{\mu}}}$
denotes the magnitude of the total dipole moment. The information about
the electric field strength and the magnitude of the dipole moment
is now incorporated into $\hat{V}_{R}$ and enters through the
radiation reorganization energy $\lambda_{R}$
introduced later.  For simplicity,  dipoles are oriented in the direction of the radiation field so that the scalar form of the dipole interaction
is adopted.   The sunlight consists of photons, so its radiation field $\hat{E}$
or $\hat{V}_R$ is treated as a photon bath,
which is fully characterized  by the \textit{energy gap correlation function} (EGCF)
\begin{equation}
C_R(t)=\langle \hat{V}_R(t)  \hat{V}_R(0)  \rangle.\label{eq:EGCF-general}
\end{equation}
The time-dependence of $\hat{V}_R(t)$ denotes the interaction picture with respect to the radiation field.
We use the over-damped  harmonic bath EGCF \cite{MukamelBook}, i.e.,  the Drude-Lorentz spectral density,
which is paramertized by  the radiation temperature $T_R=6000\;\mathrm{K}$, cutoff frequency $\gamma_R=0.1\;\mathrm{fs}-1\;\mathrm{fs}^{-1})$ and
reorganization energy $\lambda_R$  (see Appendix A).  In the weak excitation regime, the excited state density matrix is proportional to $\lambda_R$,
which can now be taken a normalization constant and needs not be specified.
As this point,  we have introduced the system Hamiltonian $\hat{H}_S$ and radiation-system interaction Hamiltonian $\hat{H}_{S-R}$,
and thus have completely defined a closed dimer system pumped by sunlight.

Light-harvesting complexes are embedded in the protein environment and therefore are open systems coupled to
phonon baths.  This coupling destroys the dynamic coherence and is described as~\cite{MayKuehnBook,Olsina2010a}
\begin{align}
\hat{H}_{S-B}=   \hat{V}_1 |\bar{e}_1\rangle \langle \bar{e}_1 |  + \hat{V}_2 |\bar{e}_2\rangle  \langle \bar{e}_2 |
\end{align}
where $\hat{V}_n$ is  the interaction strength between the molecule and its phonon bath.
Similar to the photon bath, we use the Drude-Lorentz phonon spectral density
and define the EGCF as
\begin{equation}
C_n(t)=\langle \hat{V}_n(t) \hat{V}_n(0)  \rangle,
\end{equation}
where $\hat{V}_n(t)$ is a linear function of phonon operators expressed  in the interaction picture with respect to the phonon bath.
For simplicity,  the phonon baths coupled to the two molecules share the same
parameters:  $T_1=T_2=300\;\mathrm{K}$,  $\gamma_1=\gamma_2=100\;\mathrm{cm}^{-1}$,
and $\lambda_1=\lambda_2$ in the range of $10\;\mathrm{cm}^{-1}-100\;\mathrm{cm}^{-1}$.

To calculate photon-coupling and phonon-coupling, we use the hierarchical equation of motion (HEOM), which are
developed for the Debye-Lorentz spectral density.  Details of HEOM can be found in the Appendix B.
Because of the high-temperature and extremely short coherence time of the sunlight,
the white noise description of the excitation by solar radiation is reliable, where the radiation field is treated classically.
Hence, in absence of the exciton-phonon interaction, we have derived an analytical solution based on
the Haken-Strobl model \cite{HakenStrobl1973,Cao2009,Wu2013} to describe the pumping by the sunlight.
We present this white-noise model (WNM) in the following.

\subsection{White Noise Model}

The full quantum treatment of the incoherent light in the form of HEOM is computationally  expensive.
If the coherence time of the solar radiation is shorter  than all other time-scales of the system dynamics,
including the resonance coupling, the phonon bath time-scale, and the dephasing time, which
is typical in photosynthesis, then the white-noise model (WNM) should be well applicable for the description of the radiation.
In this model, the energy gap correlation function (EGCF) is expressed in the form
\begin{equation}~\label{cr1}
C^{R}(t)=I^{R}\delta(t)\;,
\end{equation}
where $I^{R}$ is a parameter representing coupling of the electric field to the given exciton transition.
The quantum dynamics under this classical white noise is known as the Haken-Strobl model, which is exactly solvable in some cases,
including the V-shape three-level system with a trap (see \cite{Cao2009}).
We can generalize the classical noise to quantum noise by evaluating the Redfield pumping rates given as
\begin{subequations}\label{eq:WhiteNoiseSolutionRates}
\begin{align}
I_{+}^{R}= & \frac{2\gamma_{R}\lambda_{R}\varepsilon_{+}(\coth(\beta_R\varepsilon_{+}/2)-1)}{\gamma_{R}^{2}+(\varepsilon_{+}/\hbar)^{2}}\;,\\
I_{-}^{R}= & \frac{2\gamma_{R}\lambda_{R}\varepsilon_{-}(\coth(\beta_R\varepsilon_{-}/2)-1)}{\gamma_{R}^{2}+(\varepsilon_{-}/\hbar)^{2}}\;,
\end{align}
\end{subequations}
where $\varepsilon_i~(i=\pm)$ denote eigenstate energies and $\beta_R=1/(k_bT_R)$.
In the weak field regime, $I^{R}\ll k$,  the pumping creates no more than one excitation at any given time such that  $\rho_{\bar{g}\bar{g}}\approx1$.
Then, the solution to the WNM is explicitly written as
\begin{subequations}\label{eq:WhiteNoiseSolution2}
\begin{align}
\rho_{++}(t)= & \frac{\mu_{+}^{2}I_{+}^{R}}{k\mu^{2}}(1-e^{-kt})\;,\\
\rho_{--}(t)= & \frac{\mu_{-}^{2}I_{-}^{R}}{k\mu^{2}}(1-e^{-kt})\;,\\
\rho_{+-}(t)= & \frac{\mu_{+}\mu_{-}}{\mu^{2}}\frac{({I_{+}^{R}+I_{-}^{R}})/2}{k+\frac{i}{\hbar}(\varepsilon_{+}-\varepsilon_{-})}
(1-e^{-\frac{i}{\hbar}(\varepsilon_{+}-\varepsilon_{-})t-kt})\;.\label{eq:RWNM2-coherence}
\end{align}
\end{subequations}
where all parameters are given in the excitonic basis.  As long as the decay rates from the two molecules are identical and the dimer system is not coupled to any phonon bath,
the above solution retains the same functional form for arbitrary inter-site coupling $J$.
This is exactly the reason that the local and excitonic basis sets are equivalent for a closed system and the excitonic coupling $J$ does not change the physics.
The detail derivation can be obtained in the Appendix C.

\begin{figure}
\begin{center}
\centerline{\includegraphics[scale=0.3]{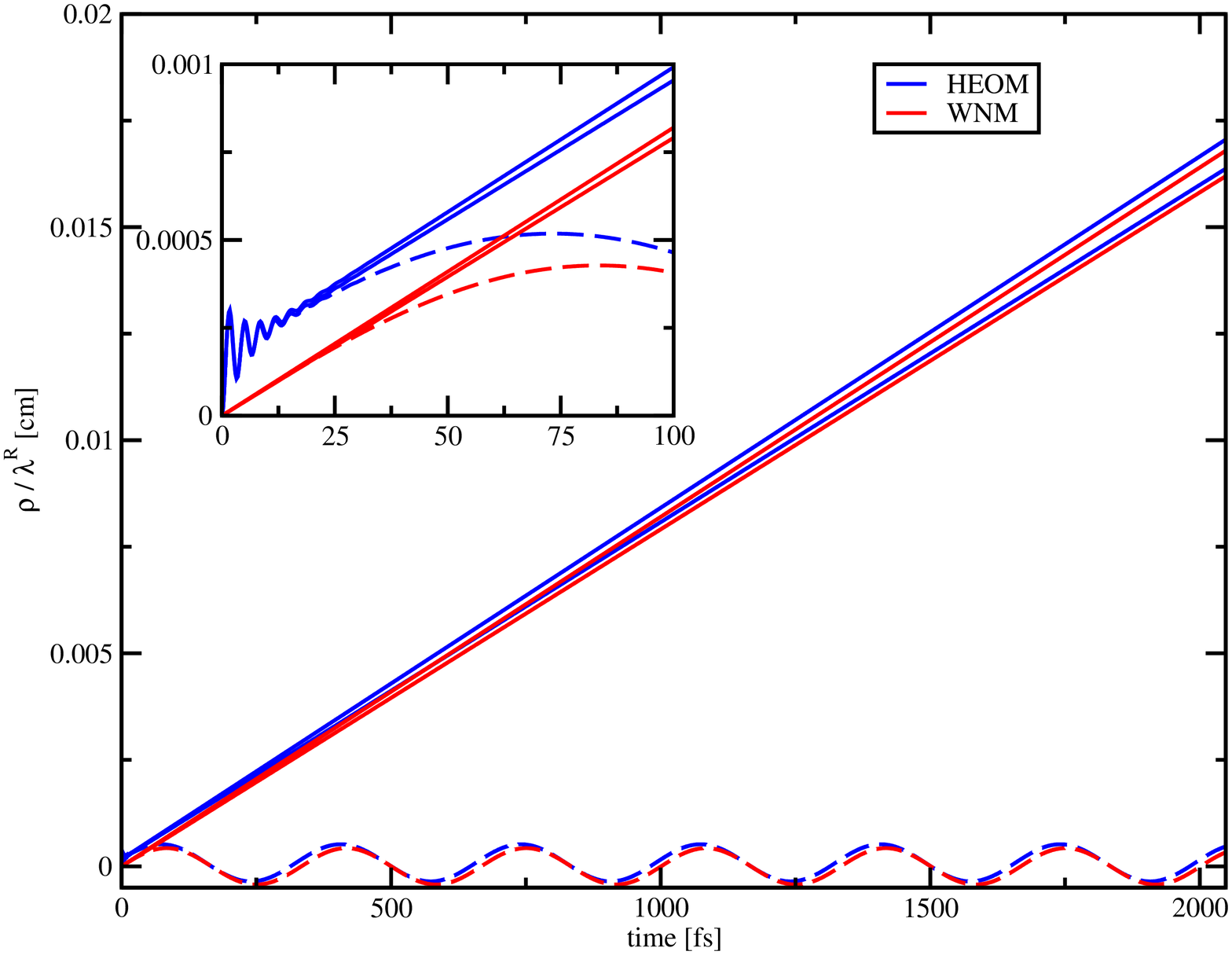}}
\caption{ Dynamics of a molecular dimer with parameters
$J=0,\epsilon_{1}=9900\;\mathrm{cm}^{-1},\epsilon_{2}=10000\;\mathrm{cm}^{-1}$
pumped by weak radiation characterized by parameters $T_{R}=6000\;\mathrm{K},\gamma_{R}=0.2\;\mathrm{fs}^{-1}$
according to the HEOM and WNM respectively.  Populations (full lines) and real
part of the coherence (dashed lines) are plotted. Apart from the transient
effects at short time shown in the inset, both models give similar results.} \label{fig:HEOM-RWNM}
\end{center}
\end{figure}

\begin{table*}
\centering
\caption{ Pumping rates $p_{+}$ and $p_{-}$ in units
of the reorganization energy according to the HEOM and WNM models
for a closed system pumped by incoherent light. Parameters of the
system are described in the caption of Fig.\ \ref{fig:HEOM-RWNM}.
The small differences may be explained  by the Markov approximation.}\label{tab:HEOM-RWNM}

\begin{tabular*}{\hsize}{@{\extracolsep{\fill}}lcr}
\hline
pumping rate & HEOM & WNM \cr
\hline
$p_{+}/\lambda_{R}$ & $8.25\cdot10^{-6}\;\mathrm{cm}\;\mathrm{fs^{-1}}$  & $8.20\cdot10^{-6}\;\;\mathrm{cm}\;\mathrm{fs^{-1}}$ \cr
$p_{-}/\lambda_{R}$ & $7.91\cdot10^{-6}\mathrm{\;\mathrm{cm}\;\mathrm{fs^{-1}}}$ & $7.91\cdot10^{-6}\;\;\mathrm{cm}\;\mathrm{fs^{-1}}$ \cr
\hline
\end{tabular*}
\end{table*}

\section{Results and Discussions}

\subsection{Excitonic Coherence in Closed Systems\label{sub:Coherence-Isolated-Systems}}

Without a distinct physical process to set the initial time $t_{0}$
as the starting point of the dynamics of natural light excitation,
we should, strictly speaking, only analyze the long time steady-states
of the density matrix. We cannot  speak about a ``precise
photon arrival time'' if there is no particular initiation process~\cite{Brumer2012a}, and thus the interpretation
of such induced dynamics is not straightforward. We can,
however, obtain practical information about the system from its dynamics.
There is, as we will show later, also
a direct link between the dynamics initiated by an ultrashort pulse
and the steady-state distribution in the weak field limit. In
this section, we present results based on the assumption that the
dynamics of the dimer under pumping by weak incoherent light starts
in the electronic ground state with no initial entanglement with the
radiation field or the phonon bath.

First, in Fig.~2 we compare the full HEOM model and the
WNM for a closed system without coupling to the phonon bath or decay to the ground state.
The choice $J=0$ is special, but the result is equally valid for any closed systems with
$J \neq 0$. The more general case of open systems will be examined later.
Both excited states have equal dipole moments with identical orientations, and the
difference in pumping rates is due to the difference in the transition frequency.
The radiation field is specified with a temperature $T_{R}=6000\;\mathrm{K}$ and cutoff
frequency $\gamma_{R}=0.2\;\mathrm{fs}^{-1}$. We work in the limit
of weak radiation electric field $\lambda_{R}\approx0$, where the
white noise model (WNM) is valid, provided the coherence time of the
radiation field is sufficiently short.  For weak fields,  both the electric field intensity
and the transition dipole strength are included in the reorganization energy of the radiation, $\lambda_{R}$,
and thus to first order the density matrix elements grow linearly in time
proportional to $\lambda_{R}$.

Fig.\ \ref{fig:HEOM-RWNM} compares  the HEOM and WNM models.
As discussed above, for a closed system, the populations grow linearly in time,  while the coherence oscillates
with a constant amplitude inversely proportional to the energy gap.
One should therefore expect more dynamical coherence to be generated
for small energy gaps. This dependence can be
easily understood in the framework of  perturbation theory,
where to leading order every molecule interacts with the field only once.
Thus  the only term originating from population excitation is $\hat{{\mu}}(\tau)|\bar{g}\rangle\langle\bar{g}|\hat{{\mu}}(\tau')$,
where $\tau$ and $\tau'$ are the times of excitation of
the right and left wave-functions and $\hat{\mu}(\tau)$ is the dipole moment operator in the Heisenberg picture.
Since the radiation field is $\delta$-correlated, $\tau'=\tau$,  we can describe the molecules
excited at times $t_{n}$\textbf{ } as an ensemble of wave-functions in the excitonic basis
\begin{equation}
|\Psi_{n}\rangle=\left[\mu_{+}|\bar{e}_{+}\rangle e^{-\frac{i}{\hbar}\epsilon_{+}(n-1)\Delta t}+\mu_{-}|\bar{e}_{-}\rangle e^{-\frac{i}{\hbar}\epsilon_{-}(n-1)\Delta t}\right]e^{i\varphi_{n}}\;,\label{eq:PSI-n-ensemble-wavefun}
\end{equation}
where $\varphi_{n}$ represents the random global phase obtained from
the incoherent light and $\Delta t$ is the discretization time step. Coupled to the same radiation field,
the excited states are coherent with each other.
We can write the total coherence averaged over all random phases as
\begin{align}
\rho_{+-}& = \sum_{n=1}^{t/\Delta t}\frac{1}{(2\pi)^{n}}\int_{0}^{2\pi}  d\varphi_{1}\dots\int_{0}^{2\pi}d\varphi_{n}\langle\bar{e}_{+}|\Psi_{n}\rangle\langle\Psi_{n}|\bar{e}_{-}\rangle \nonumber \\
 &= \sum_{n=1}^{t/\Delta t}\mu_{+}\mu^*_{-}e^{-\frac{i}{\hbar}(\epsilon_{+}-\epsilon_{-})(n-1)\Delta t}
= i \hbar \mu_+ \mu_-^*  { e^{-\frac{i}{\hbar} \epsilon t} -1 \over \epsilon}.
\end{align}
which is inversely proportional to the energy gap $\epsilon=\epsilon_+-\epsilon_-$.
This is a general expression for arbitrary coupling $J$.
The same result can be obtained from the decoherence theory, where one simply replace
the random phases in Eq.\ (\ref{eq:PSI-n-ensemble-wavefun}) with the states
entangled with the radiation field and the integrals over random phases
are then replaced by the trace over the radiation field.  A similar result was also obtained independently
by the Brumer group.\cite{Brumer2014}

\begin{figure}
\begin{center}
\centerline{\includegraphics[scale=0.3]{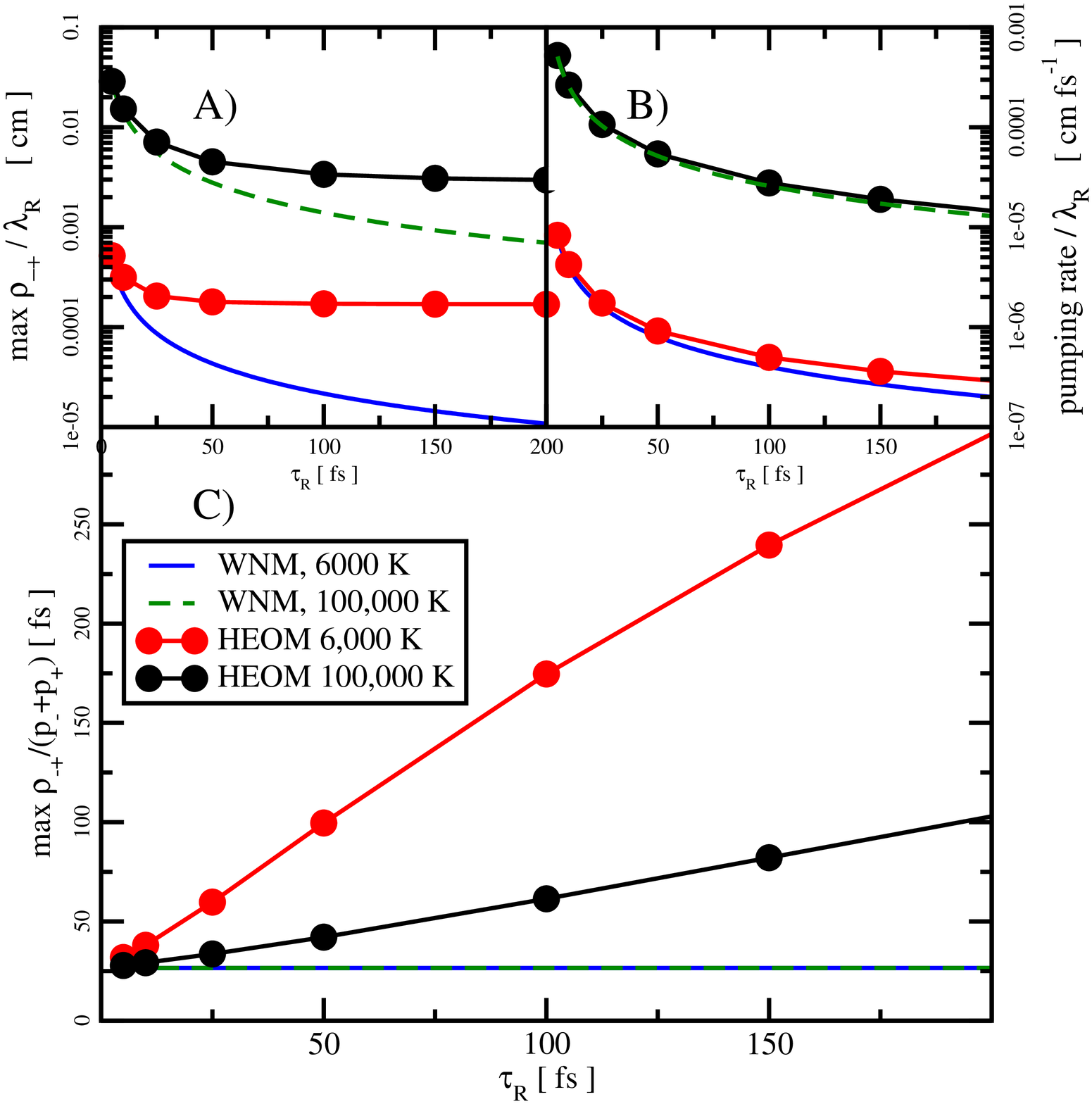}}
\caption{ Comparison between the HEOM and WNM
models in the weak radiation regime  for gradually increasing
coherence time of the sunlight at the radiation temperature of $T_{R}=6000\;\mathrm{K}$
or $T_{R}=100,000\;\mathrm{K}$. Fig.\ (a) shows the maximal amplitude
of the steady state coherence $\rho_{-+}(t)$ extracted at long times with $\gamma_R=0.2$~fs$^{-1}$. Fig.\ (b)
shows the dependence of the pumping rate $p_{+}$ on the coherence time of the radiation field, $\tau_{\mathrm{R}}$.
The pumping rate $p_{2}$ is almost the same as $p_{-}$ and is not
plotted. Fig.\ (c) shows the normalized coherence $\max\{\rho_{-+}(t)\}/{(p_{+}+p_{-})}$,
which is independent of the radiation intensity for weak fields.
The other parameters are given by $J=0$, $\epsilon_1=9900$~cm$^{-1}$ and $\epsilon_2=10000$~cm$^{-1}$.
} \label{fig:CoherenceProperties}
\end{center}
\end{figure}

As can be seen in the insert of Fig.\ \ref{fig:HEOM-RWNM}, the difference between
the HEOM and WNM models are the small transient dynamics upon excitation,
resulting in a slight offset in the populations and a phase shift in the coherence.
Otherwise, the overall agreement between the two models is excellent. Table\ 1 
quantifies the pumping rates predicted by both models. In the weak field regime, the HEOM result
should reproduce the Redfield rate.

Further, we study the dependence
on the coherence time of the radiation, $\tau_{\mathrm{R}}=1/\gamma_{R}$.
The agreement between HEOM and WNM deteriorates with
the increase of the light coherence time, $\tau_{\rm R}$.
To quantify this difference, we use  two quantities that can easily be extracted
from the exciton dynamics: The maximum of the oscillating coherence $\rho_{+-}$ (or $\rho_{-+}$),
taken after the transient dynamics,
and the pumping rates $p_{+}$ and $p_{-}$, defined
as the derivatives of the linearly growing populations $\rho_{++}(t)$
and $\rho_{--}(t)$.  We compare the two quantities for two radiation temperatures of
$T_{R}=6000\;\mathrm{K}$ and $T_{R}=100,000\;\mathrm{K}$. The latter
conveniently represents the classical limit  at high temperatures.
As can be seen in Fig.\ \ref{fig:CoherenceProperties}, all models
match well at short coherence time, $\tau_{\mathrm{light}}=5\;\mathrm{fs}$, but with
a gradual increase of the coherence time, the WNM model underestimates
the amount of coherence present in the system, particularly for the
temperature $6000\;\mathrm{K}$.  While the WNM captures the pumping
rates well, the predicted coherence term  converges to zero instead of a constant
for long coherence time $\tau_{\mathrm{light}}$.

\begin{figure}
\begin{center}
\centerline{\includegraphics[width=0.9\columnwidth]{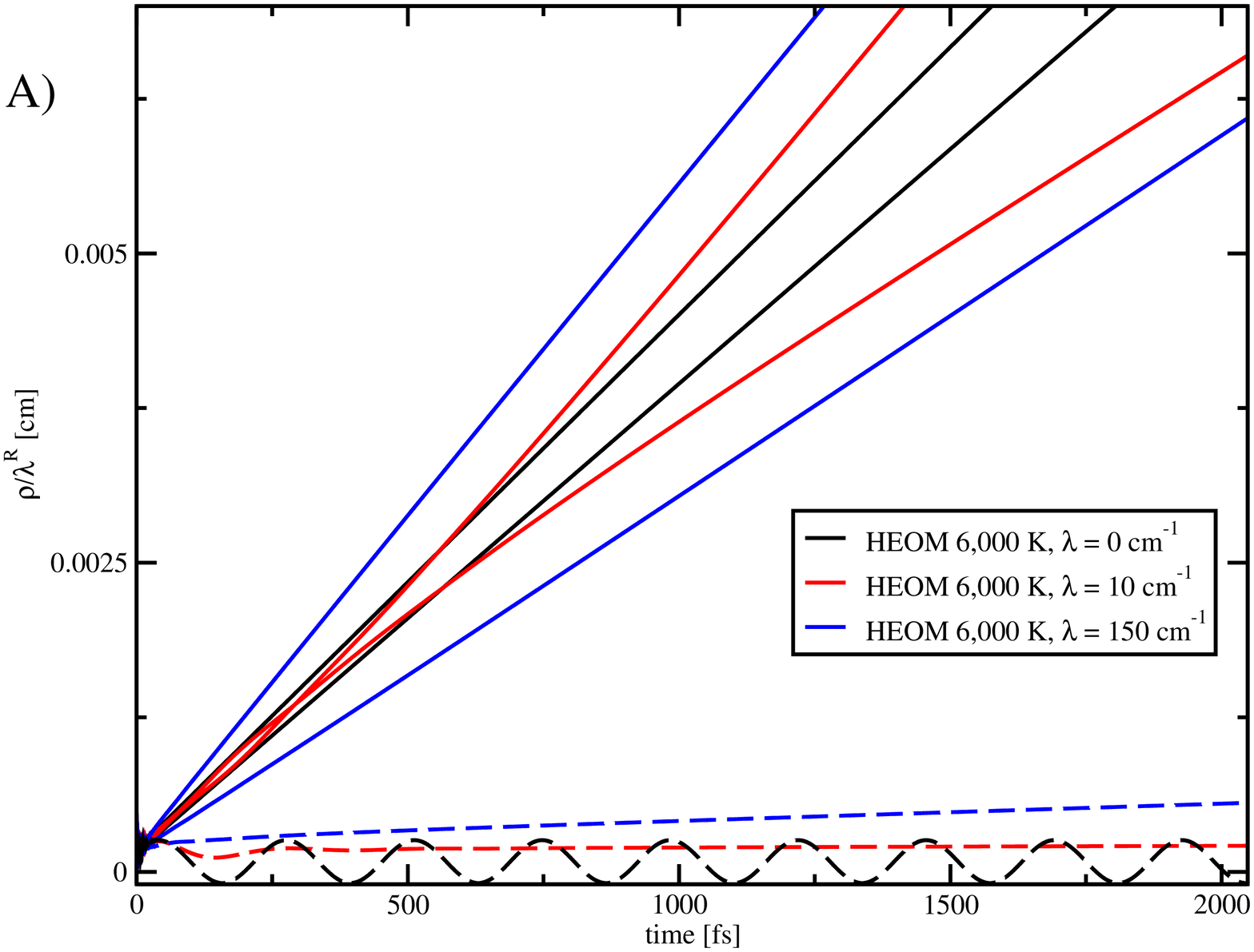}}

\centerline{\includegraphics[width=0.9\columnwidth]{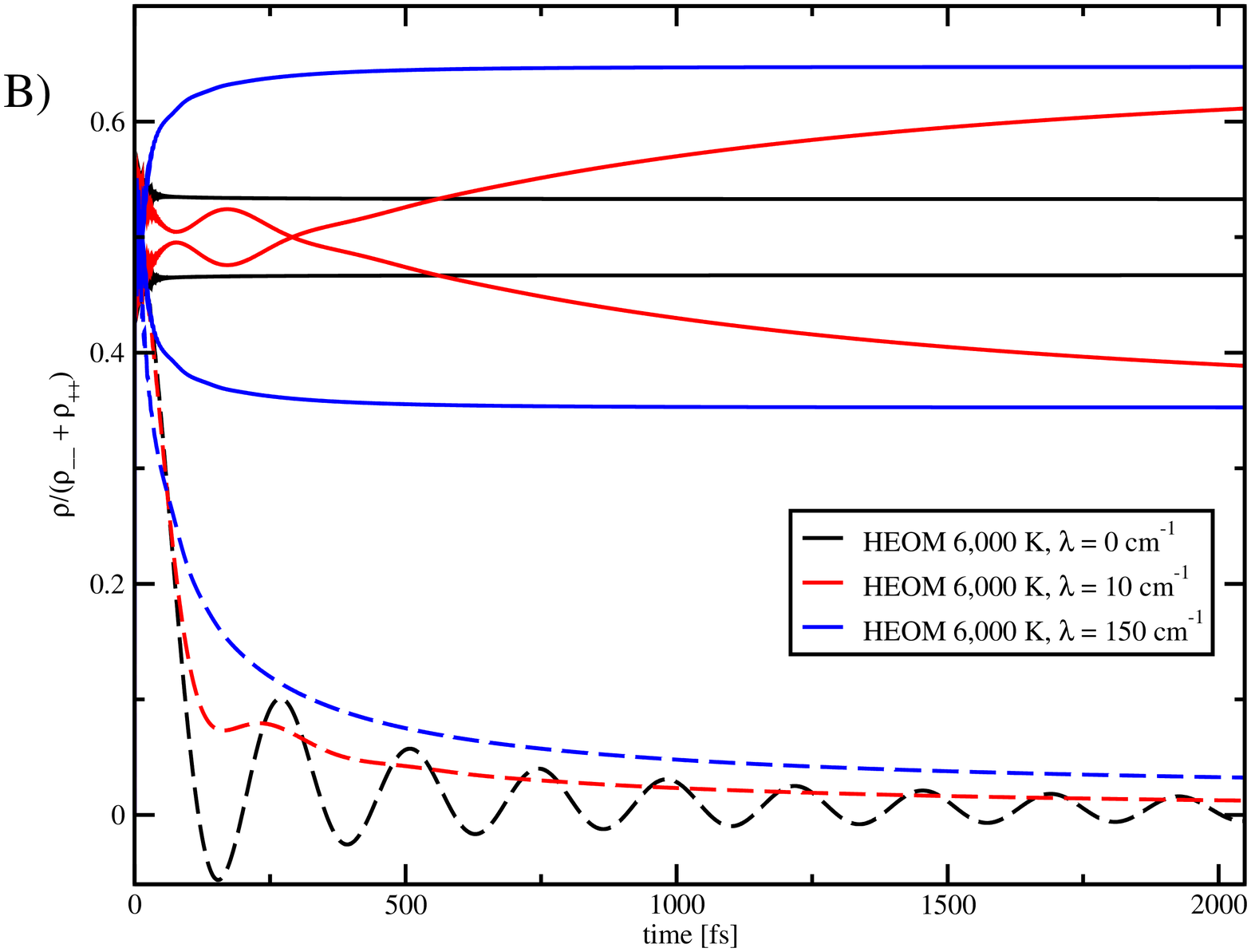}}
\caption{ Dynamics of an open dimer system described in Results and Discussion section. 
 Fig.\ (a) shows the population
and the real part of the coherence for $\lambda=0$, $\lambda=10\;\mathrm{cm}^{-1}$
and $\lambda=150\;\mathrm{cm}^{-1}$ in black, red and blue lines.
Fig.\ (b) shows the population and the coherence normalized by the
sum of populations of single-excited states, i.e.,
the relative coherence within the manifold of single-excitedstates.
} \label{fig:OpenSystems}
\end{center}
\end{figure}

\begin{figure}
\begin{center}
\centerline{\includegraphics[width=0.85\columnwidth]{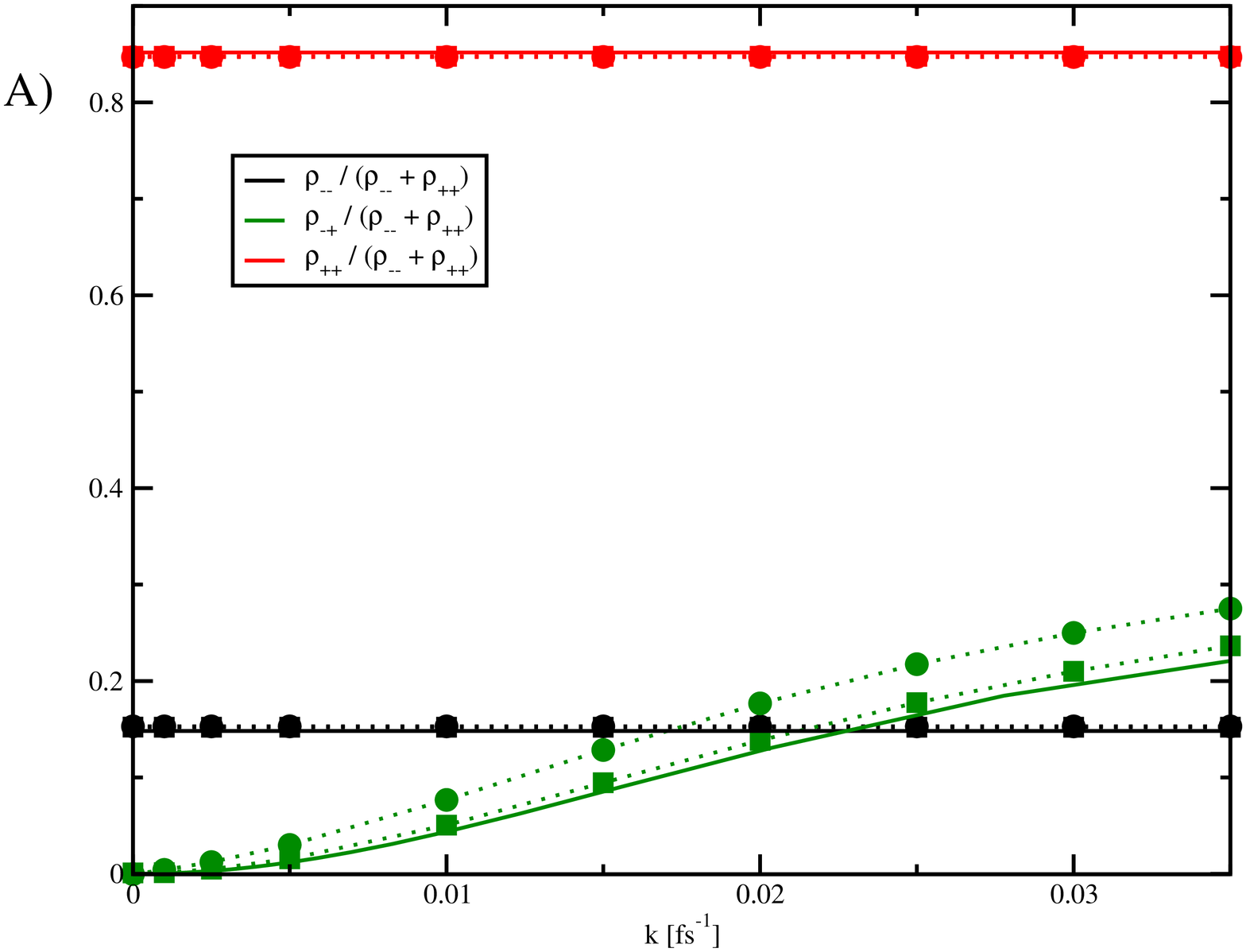}}

\centerline{\includegraphics[width=0.85\columnwidth]{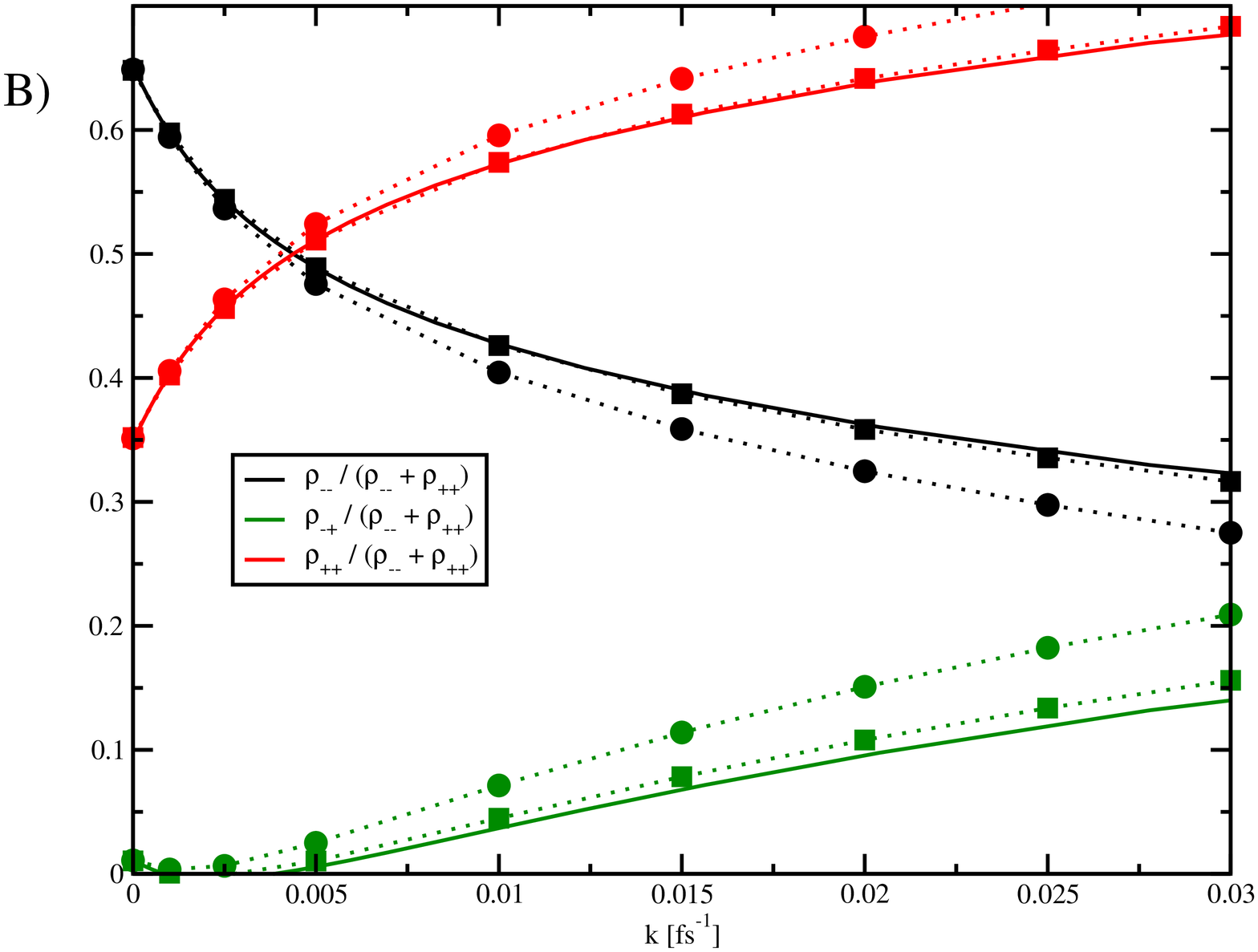}}
\caption{\label{fig:Equilibria} Dependence of the dimer system's non-equilibrium steady-state
on the decay rate $k$. The system parameters
are $J=50\;\mathrm{cm}^{-1},\epsilon_{1}=9900\;\mathrm{cm}^{-1},\epsilon_{2}=10000\;\mathrm{cm}^{-1}$, and
the effective temperature of the radiation field is $T_{R}=6000\;\mathrm{K}$. The
HEOM results for $\gamma_{R}=0.1\;\mathrm{fs}^{-1}$  and $\gamma_{R}=1\;\mathrm{fs}^{-1}$
are shown with circle symbols and square symbols, respectively.
The theoretical WNM result given by Eq.\ (\ref{eq:EnsembleWithGamma}) is shown in full
lines. Fig.\ (a) shows  the results without coupling to a phonon bath.  Fig.\ (b) shows
the results with coupling to a phonon bath for which $\lambda_{1}=\lambda_{2}=100\;\mathrm{cm}^{-1},\gamma_{1}=\gamma_{2}=100\;\mathrm{fs}$
and temperature $T_{1}=T_{2}=300\;\mathrm{K}$. Changing $k$ scans
the system dynamics: for fast $k$, the system approaches a pure state
$\hat{\mu}|\bar{g}\rangle\langle\bar{g}|\hat{\mu}$; for slow $k$,
the system steady-state is given by the phonon bath thermal state
Eq.\ (\ref{eq:rho-eq}) without pumping, including the small amount
of coherence due to bath redefinition of the basis. The steady states
between these two extremes are closely related to the dynamics
under no light pumping initiated with a $\delta$-pulse through Eq.\ (\ref{eq:EnsembleWithGamma}).}
\end{center}
\end{figure}

\subsection{Excitonic Coherence in Open Systems\label{sub:Coherence-in-Open-Systems}}

The systems relevant in the primary processes of photosynthesis
are open systems, so here we investigate the excitonic dynamics
of open systems under pumping by incoherent light in the HEOM model.
Unlike  in the previous discussion of closed systems, the choice $J=0$ would be a special case
and is no longer assumed; instead,  $J=50$~cm$^{-1}$ is used.
The radiation field has the temperature of $T_{R}=6000\;\mathrm{K}$
and a coherence time of $\tau_{\mathrm{light}}=10\;\mathrm{fs}$. The coupling
to the electromagnetic field is assumed to be weak such that  the density matrix can
be normalized by the reorganization energy of the radiation field,  $\lambda_{R}$.
The dipole moments are parallel and have relative strengths of $\mu_{1}/\mu=0.60$ and
$\mu_{2}/\mu=1.28$ in the local basis. The phonon baths are
uncorrelated between the two sites, and have the same reorganization energy
$\lambda=\lambda_{1}=\lambda_{2}$, temperature $T_{1}=T_{2}=300\;\mathrm{K}$,
and cutoff frequency $\gamma_1=\gamma_2=100\;\mathrm{cm}^{-1}$.
Fig.\ \ref{fig:OpenSystems} gives the time evolution of the exciton
populations and coherence for three values of the phonon bath reorganization energies,
$\lambda=0$ (closed system), $\lambda=10\;\mathrm{cm}^{-1}$ and
$\lambda=150\;\mathrm{cm}^{-1}$, respectively.  For comparison,
 the same dynamics is normalized by the sum of excited state populations, which better demonstrates
the relative coherence between the excited states.

We can make several observations about the exciton dynamics. For the
closed system, the ratio of the populations is given solely by the
couplings of the states to the electromagnetic field. In the presence
of the phonon bath, this ratio is given by the coupling to the radiation at short times,
which corresponds to the state $\hat{{\mu}}|\bar{g}\rangle\langle\bar{g}|\hat{{\mu}}$,
and gradually approaches the reduced thermal equilibrium, i.e., the static coherence,
which is  determined by the coupling to  the phonon bath, $\rho_{eq}=\lim_{t\rightarrow\infty} \rho(t)$.
To understand the effect, we can again invoke the ensemble picture discussed
in the previous section and the description of second-order interactions
with the radiation field.    There is always a portion of
molecules that were excited only recently and their state is very
close to the pure quantum state $\hat{{\mu}}|\bar{g}\rangle\langle\bar{g}|\hat{{\mu}}$
with population ratios given by their coupling to the radiation. Every
molecule, however, reaches the thermal equilibrium state determined by the phonon
bath after a certain time, and the portion of such molecules
grows linearly, which explains the behavior of the trace-normalized excited
state manifold (Fig.\ \ref{fig:OpenSystems}(b) in the short and long
time limits.
The decay  time of the coherence depends  weakly on
the phonon bath reorganization energy.  However, in comparison to the
closed system, even weak coupling
to the phonon bath damps the oscillations observed in the coherence
and give rise to a continuously decaying non-oscillatory coherence.
At long times, the coherence does not decay to zero, as in an isolated system, but
approaches a constant plateau.
The non-vanishing plateau value of the coherence arises from the
entanglement with the phonon bath, and this static coherence cannot
be captured by the WNM~\cite{Moix2012}.

\subsection{Steady-State Distribution with Population Decay\label{sub:Discussion-Equilibrium-properties}}

Whether one should describe a sample of continuously excited molecules
as an ensemble composed of molecules undergoing coherent dynamics,
or adopt the point of view of decoherence theory, where every individually-excited
molecule loses coherence through its entanglement with the radiation and
phonon baths, is still an open question.  Here, we establish a relationship between
the non-equilibrium steady-state of the exciton system
in the presence of a decay rate $k$
and  the dynamics initiated by a coherent short pulse,
which is usually probed by ultrafast nonlinear spectroscopy experiments.

Following the ensemble description of the dephasing dynamics, Eq.\ (\ref{eq:PSI-n-ensemble-wavefun}),
we can write the steady-state of the excited molecules as
\begin{equation}
\rho_{\mathrm{steady}}\approx\int_{0}^{\infty}d\tau\; e^{-k\tau}\mathcal{U}(\tau)\hat{{\mu}}|\bar{g}\rangle\langle\bar{g}|\hat{{\mu}}\;,\label{eq:EnsembleWithGamma}
\end{equation}
where $\mathcal{U}(t)$ is the radiation free evolution super-operator,
representing the  dissipative dynamics resulting from the coupling to
 the phonon bath alone. Eq.\ (\ref{eq:EnsembleWithGamma}) is exact
for the $\delta$-correlated sunlight for which $\gamma_{R}\to\infty$. In
other words, for strongly incoherent light, we can decompose the long
time steady-state of the system into an ensemble of molecules excited
by a $\delta$-pulse at different times, weighted by the survival probability
determined by the relaxation to the ground state. Whether
or not such a decomposition is truly physical, Eq.\ (\ref{eq:EnsembleWithGamma})
shows a direct relationship  between ultrafast coherent dynamics and
the steady-state under pumping by incoherent light.
The decay ate $k$ is a natural timescale to be compared with the lifetime of
electronic coherence. The coherent dynamics
only manifests in the steady state
if the timescale for the decay to the ground state is comparable to the lifetime of
dynamic coherence.

To demonstrate this connection, we study the dependence of the steady-state
 on the decay rate $k$. Calculations are performed by
the HEOM on a molecular dimer with parameters $J=50\;\mathrm{cm}^{-1}$,
$\epsilon_{1}=9900\;\mathrm{cm}^{-1}$, $\epsilon_{2}=10000\;\mathrm{cm}^{-1}$,
pumped by weak light with temperature $T=6000\;\mathrm{K}$ and $\gamma_{R}=0.1\;\mathrm{fs}^{-1}$.
Two cases  are studied: a closed system with no phonon bath and an open system with
uncorrelated phonon baths.
Fig.\ \ref{fig:Equilibria} shows the dependence of the system steady-state
on the decay rate $k$.   Here, the trace of the density matrix on the excited
state manifold is normalized. For a closed system (Fig.\ \ref{fig:Equilibria}(a),
we observe a gradual transition from the pure state $\hat{{\mu}}|\bar{g}\rangle\langle\bar{g}|\hat{{\mu}}$
for fast $k$, which corresponds to a $\delta$-pulse excitation,
to the diagonal part of $\hat{{\mu}}|\bar{g}\rangle\langle\bar{g}|\hat{{\mu}}$
for slow $k$.
In the presence  of a phonon bath (Fig.\ \ref{fig:Equilibria}(b)), the same transition happens
between the state $\hat{{\mu}}|\bar{g}\rangle\langle\bar{g}|\hat{{\mu}}$
for fast $k$ and the thermal equilibrium state of the system under no pumping
\begin{equation}
\rho_{\mathrm{eq}}=\lim\limits _{t\to\infty}\mathcal{U}(t)\hat{{\mu}}|\bar{g}\rangle\langle\bar{g}|\hat{{\mu}}
\label{eq:rho-eq}
\end{equation}
for slow $k$.  The reduced equilibrium distribution of the system is not canonical because of the coupling to the phonon bath
and can be obtained directly by stochastic path integral simulations or polaron transformation~\cite{Moix2012}.
The steady state dependence shown in Fig.\ \ref{fig:Equilibria}(b)
is more complicated in comparison with the closed system shown in Fig.\ \ref{fig:Equilibria}(a).
 If  we compare the steady-state dependence with the result given by Eq.\ (\ref{eq:EnsembleWithGamma})
for perfectly white noise, there are detectable differences from the HEOM
result with $\tau_{\mathrm{R}}=10\;\mathrm{fs}$, but these differences
disappear for $\tau_{\mathrm{R}}=1\;\mathrm{fs}$.

Although it is difficult to tune the decay rate $k$ experimentally,
there is a close connection between the
dynamics initiated by a $\delta$-pulse and the dynamics induced
by natural sunlight.  Since the reduced density matrix contains all the
information about the measurements performed on the
system, the excitonic  coherence can have an effect on the energy
transfer only if the light-induced coherent dynamics affects the steady
state through Eq.\ (\ref{eq:EnsembleWithGamma}). For this reason,
if the excitonic coherence probed by 2DES spectroscopy plays a significant
role in  energy transfer, it will be relevant only in situations,
where the decoherence lifetime is sufficiently long compared to the decay rate
in the system. Although the decoherence lifetime in photosynthetic systems
can be as high as several picoseconds \cite{Mercer2009a,Panitchayangkoon2011a},
this time-scale may not exceed the decay lifetime to the reaction
center, which is approximately 50 ps  \cite{Sundstrom1986, Timpmann1991, Zhang1992} in many species of photosynthetic bacteria.
For this reason, enhancement of the energy transfer by the dynamical
coherence induced by sunlight may not be a dominant effect.  In contrast,
 the spatial, static coherence  due to quantum delocalization has a persistent effect
 on energy transfer efficiency (e.g. LH2) and robustness (e.g. FMO), and this effect is
 independent of light-induced coherence.

\section{CONCLUSIONS}
Our results demonstrate that the incoherent nature of sunlight excitation does not exclude transient coherence in the light-induced exciton
dynamics but this dynamical coherence may not play a dominant role in light-harvesting energy transfer which is governed by static coherence resulting from the coupling to phonons.
\begin{itemize}
\item
For a realistic light-harvesting system, the excitonic coherence is dynamical at short times, but becomes static at long times,
which corresponds to the delocalization of states.
The amount of dynamical coherence of a closed system is inversely proportional to the excitonic energy gap,
and  depends the coupling to the phonon bath.
\item
The decay to the ground state establishes a non-equilibrium steady state and defines an observation window for the induced exciton dynamics.
Under the influence of the phonon bath, as the decay time increases,
 the steady state population distribution changes from photon-induced
to phonon-induced, and the steady state coherence changes from dynamical to static.
In the fast decay rate limit, the amount of steady coherence corresponds to the dynamical coherence and increases with the decay rate.
whereas the slow-decay limit of the coherence is exactly the reduced equilibrium distribution.
\item
The contribution of the dynamical coherence to the system steady-state depends critically on the ratio of the lifetime of the dynamical coherence  and the decay rate to the ground state.
For photosynthetic light-harvesting systems,  the light-induced dynamics lasts for hundreds of femtoseconds, whereas the observed energy trapping occurs on tens of picoseconds; therefore  the light-induced coherence
is dissipated on the trapping time-scale and is generally not a major factor consideration in efficiency light-harvesting energy transfer.
\item
Theoretically, the proposed white-noise model (WNM) with Redfield rates provides a reliable description of the excitation dynamics.  As  a result, the short coherence time of sunlight
enables us to establish a simple connection between the dynamics excited by an ultra-short laser pulse as probed by 2DES spectroscopy and the steady-state of the excitonic system under pumping.
\end{itemize}

\section*{Acknowledgments}

This work was supported by the National Science Foundation (Grant CHE-1112825) and the DOE.  Arend Dijkstra and Jianshu Cao were supported by the Center of Excitonics,
an Energy Frontier Research Center funded by the US Department of Energy, Office of Science, Office of Basic Energy Sciences
under Award DE-SC0001088. Jan Ol\v{s}ina was supported by the Karel Urb\'{a}nek Fund.
Jianshu Cao thanks Prof. Graham Fleming and Prof. Paul Brumer  for helpful discussions
and  Dr. Dong Hui of the Fleming group for sharing a derivation similar to the Appendix C.

\appendix

\section{Quantum description of the sunlight\label{sub:Quantum-light}}
To ensure generality of the results, we use a fully quantum description of the radiation.
The radiation bath and system-bath interactions are in the Schr\"odinger picture defined as
\cite{GreinerBook}
\begin{align}
\hat{H}_{R}= & \;\sum_{\sigma\mathbf{k}}\hbar\omega_{\mathbf{k}}\left(\hat{b}_{\sigma\mathbf{k}}^{\dagger}\hat{b}_{\sigma\mathbf{k}}+\frac{1}{2}\right)\;,\\
\hat{H}_{S-R}= & \;-\sum_{m\in\{x,y,z\}}\hat{E}_{m}{\otimes}\hat{{\mu}}_{m}\;.\label{eq:RadiationSBC-MuE}
\end{align}
They are expressed in terms of photon creation (annihilation) operators
$\hat{b}_{\sigma\mathbf{k}}^{\dagger}~(\hat{b}_{\sigma\mathbf{k}})$ for different
wave-vectors $\mathbf{k}$, polarizations $\sigma$ and frequencies
$\omega_{\mathbf{k}}$. The interaction term in the dipole approximation
couples the system and radiation field. The coupling to the system
is through the $m$-th spatial component of the total dipole-moment
operator $\hat{\mu}_{m}$.
The radiation field part is described by
\begin{equation}\label{eq:EfieldQuant}
\hat{E}_{m}=i\sum_{\mathbf{k}}N_{\mathbf{k}}\boldsymbol{\varepsilon}^{\mathbf{k}m}\left(\hat{b}_{m\mathbf{k}}-\hat{b}_{m\mathbf{k}}^{\dagger}\right)\;.
\end{equation}
 The polarization vectors $\boldsymbol{\varepsilon}^{\mathbf{k}m}$
are orthogonal to $\mathbf{k}$ and to each other.
The field is quantized in a box of a size $L$ and
the limit $L\to\infty$ is to be performed at the end of our calculations
\cite{GreinerBook}. In SI units, this corresponds to normalization constants
\begin{equation}
N_{\mathbf{k}}=\sqrt{\frac{\hbar\omega_{\mathbf{k}}}{2L^{3}\epsilon_{0}}}\;,
\end{equation}
where $\epsilon_{0}$ denotes the vacuum permittivity.

In order to express the equations in an unified way, we rewrite Eq.\ (\ref{eq:RadiationSBC-MuE}) as
\begin{equation}
\hat{H}_{S-R}=\sum_{m\in\{x,y,z\}}\hat{V}_{R_{m}}{\otimes}\hat{K}_{R_{m}}
\end{equation}
where $\hat{K}_{R_{m}}=\hat{\mu}_{m}/\mu_{m}$ and $\hat{V}_{R_{m}}=-\mu_{m}\hat{E}_{m}$.
The constant $\mu_{m}=\frac{1}{2}\sqrt{\mathrm{Tr}\,\hat{\mu}_{m}\cdot\hat{\mu}_{m}}$
denotes the magnitude of the $m$-th component of the dipole moment. The information about
the electric field strength and the magnitude of the dipole moment
is kept in the bath operators $\hat{V}_{R_{m}}$ and enters through the
radiation reorganization energy $\lambda_{R}$%
, while the system operators $\hat{K}_{R_{m}}$
hold the dimensionless matrix structure of the dipole moments. In
order to lower the computational cost, we choose both molecules to
have parallel dipole moments, which allows us to use only one bath
($\hat{K}_{R},\hat{V}_{R}$) and avoid the orientational averaging.
However, the essential physics will not be changed.
We drop the coordinate index $m$ for the total dipole moment
operator $\hat{\mu}$ and we assume that it is taken along the
axis of the dipole moment. In the manuscript, we use quantities $\mu$ and $\hat{E}$ instead of $\mu_m$ and $\hat{E}_m$.

\section{Hierarchical Equations of Motion\label{sub:Hierarchical-Equations-of}}

In order to describe the solar light  and phonon baths quantum mechanically, we use
the Hierarchical Equations of Motion as~\cite{Tanimura1989a}
\begin{align}
\dot{\hat{\rho}}^{n_{\alpha k}}(t) & =-\left(i\hat{H}_{S}^{\times}+\sum_{\alpha}\sum_{k=0}^{M}n_{\alpha k}\nu_{\alpha k}\right)\hat{\rho}^{n_{\alpha k}}(t)\nonumber \\
 & -\sum_{\alpha}\left(\frac{2\lambda_{\alpha}}{\beta_{\alpha}\gamma_{\alpha}}-i\lambda_{\alpha}-\sum_{k=0}^{M}\frac{c_{\alpha k}}{\nu_{\alpha k}}\right)\hat{K}_{\alpha}^{\times}\hat{K}_{\alpha}^{\times}\hat{\rho}^{n_{\alpha k}}(t)\nonumber \\
 & -i\sum_{\alpha}\sum_{k=0}^{M}\hat{K}_{\alpha}^{\times}\hat{\rho}^{n_{\alpha k}^{+}}(t)\nonumber \\
 & -i\sum_{\alpha}\sum_{k=0}^{M}n_{\alpha k}\left(c_{\alpha k}\hat{K}_{\alpha}\hat{\rho}^{n_{\alpha k}^{-}}(t)-c_{\alpha k}^{*}\hat{\rho}^{n_{\alpha k}^{-}}(t)\hat{K}_{\alpha}\right)\label{eq:HEOM-Full}
\end{align}
written in the standard notation~\cite{Dijkstra2010}.  HEOM introduces
an infinite set of operators $\hat{\rho}^{n_{\alpha k}}(t)$ numbered by
a multi-index $n_{\alpha k}$ to represent entanglement with the bath
and the bath memory effects. The index $\alpha\in\{R,1,2\}$ denotes
all baths present, both radiation and phonon ones, while the index
$k$ denotes the Matsubara frequencies $\nu_{\alpha k}=2\pi k/\beta_{\alpha}$
included up to some maximum frequency $M$. For convenience, we define
$\nu_{\alpha0}=\gamma_{\alpha}$.  The standard definition of inverse temperature
$\beta_{\alpha}=1/k_{B}/T_{\alpha}$ is used. Each of the integer
numbers $n_{\alpha k}$ can attain values from 0 to infinity, but
in praxis they are truncated after a sufficient number of tiers $\Theta_{\mathrm{max}}$.
Only operators $\hat{\rho}^{n_{\alpha k}}(t)$ for which $\sum_{\alpha}\sum_{k=0}^{M}n_{\alpha k}\leq\Theta_{\mathrm{max}}$
are taken into account. We use abbreviations $n_{\alpha k}^{+}=n_{\alpha k}+1$,
$n_{\alpha k}^{-}=n_{\alpha k}-1$,
\begin{align}
c_{\alpha k} & =\frac{4\lambda_{\alpha}\gamma_{k}}{\beta_{\alpha}}\frac{\nu_{\alpha k}}{\nu_{\alpha k}^{2}-\gamma_{\alpha}^{2}}\;,\\
c_{\alpha0} & =\lambda_{\alpha}\gamma_{\alpha}(\cot(\beta_{\alpha}\gamma_{\alpha}\hbar/2)-i)
\end{align}
 to make the notation more compact. The physical density matrix is
the operator $\hat{\rho}^{n_{\alpha k}}(t)$ for which all $n_{\alpha k}=0$.
The symbol $\hat{A}^{\times}$ denotes Liouville space operator (superoperator)
defined by its action on a (Hilbert space) operator $\bullet$ as
$\hat{A}^{\times}\bullet=[\hat{A},\bullet]$ for a given operator $\hat{A}$.

The equations (\ref{eq:HEOM-Full}) assume a Drude-Lorentz bath with
EGCF  in a form
\begin{align}
C^{\alpha}(t)= & \hbar c_{\alpha0}\exp(-\gamma_{\alpha}t)+\sum_{k=1}^{\infty}c_{\alpha k}\exp\left(-\nu_{\alpha k}t\right)\;,\label{eq:EGCF-overdamped}
\end{align}
where $\alpha\in\{R,1,2\}$.

The computational cost of the full form of the equations (\ref{eq:HEOM-Full})
is extremely high, especially at low temperature.
To deal with this difficulty,  a hybrid version, i.e., the stochastic-HEOM, has been developed to
simulate low-temperature dynamics, and various high-temperature approximations have
been successfully used \cite{Ishizaki2009a,Kreisbeck2011a,Moix2013},
effectively reducing the number of included Matsubara frequencies
to zero ($M=0$) or approximating the first Matsubara frequency without
further computational costs \cite{Ishizaki2009a}. These approximations
need two conditions to be met: $\beta\hbar\gamma\ll1$ and $\beta\Delta\ll1$,
where $\Delta$ is a characteristic energy gap of the system. In order
to describe incoherent natural light, we use a bath with a temperature
$T_{R}=6000\;\mathrm{K}$, coherence time $\gamma_{R}\approx0.1\;\mathrm{fs}^{-1}$
and $\Delta=10,000\;\mathrm{cm}^{-1}$, which gives us $\beta_{R}\hbar\gamma_{R}\approx0.13$
and $\beta_{R}\Delta\approx2.40$. While the first criterion is well
satisfied in our calculation, the second leads to incorrect results
and Matsubara frequencies up to the second one need to be included.

In part of presented calculations, we use HEOM (\ref{eq:HEOM-Full})
together with a decay rate $k$ from the excited state manifold to
the ground state. Is such a case, the rate is applied to all operators
$\hat{\rho}^{n_{\alpha k}}(t)$ from Eq. (\ref{eq:HEOM-Full})\begin{subequations}\label{eq:HEOM_with_gamma}
\begin{align}
\langle\bar{e}_{u}|\dot{\hat{\rho}}^{n_{\alpha k}}(t)|\bar{e}_{v}\rangle= & \;\langle\bar{e}_{u}|\dot{\hat{\rho}}_{\mathrm{HEOM}}^{n_{\alpha k}}(t)-k\hat{\rho}^{n_{\alpha k}}(t)|\bar{e}_{v}\rangle\;,\\
\langle g|\dot{\hat{\rho}}^{n_{\alpha k}}(t)|g\rangle= & \;\langle g|\dot{\hat{\rho}}_{\mathrm{HEOM}}^{n_{\alpha k}}(t)|g\rangle\nonumber \\
 & +k\sum_{u}\langle\bar{e}_{u}|\hat{\rho}^{n_{\alpha k}}(t)|\bar{e}_{u}\rangle\;.
\end{align}
\end{subequations} Here, $\dot{\hat{\rho}}_{\mathrm{HEOM}}^{n_{\alpha k}}(t)$
denotes the reduced density matrix derivative calculated from the
Eq.~(\ref{eq:HEOM-Full}). We refer to this set of equations as the HEOM
with rate.


\section{Stochastic wavefunction solution: White noise model}

In this paper, a ground state and two single-excited states of a molecular dimer are modeled as a three-level system, given by
\begin{eqnarray}~\label{hs1}
\hat{H}_S=\sum_{v=1,2}\epsilon_v|\bar{e}_{v}{\rangle}{\langle}\bar{e}_v|+\epsilon_g|\bar{g}{\rangle}{\langle}\bar{g}|+J(|\bar{e}_1{\rangle}{\langle}\bar{e}_2|+H.c.),\nonumber\\
\end{eqnarray}
where $\epsilon_v~(v=1,2,g)$ is the site energy of the exciton for the local state $|\bar{e}_v{\rangle}$ and the ground state $|\bar{g}{\rangle}$,
and $J$ is the resonance coupling between states $|\bar{e}_1{\rangle}$
and $|\bar{e}_2{\rangle}$. Without loss of generality,
we set the ground state energy $\epsilon_g=0$ as the energy reference.

Under excitation by natural sunlight, the system interacts with the radiation field via the dipole coupling as
\begin{eqnarray}~\label{hs2}
\hat{H}_{S-R}=\hat{V}_R{\otimes}\hat{K}_R=-\hat{E}{\otimes}\sum_{v=1,2}[\mu_v|\bar{e}_v{\rangle}{\langle}\bar{g}|+H.c.],
\end{eqnarray}
where the exciton dipole moment is described as  $\hat{K}_R=\sum_{v=1,2}(\frac{\mu_v}{\mu}|\bar{e}_v{\rangle}{\langle}\bar{g}|+H.c.)$,
with the magnitude of the collective dipole moment defined as $\mu=\sqrt{\mu^2_1+\mu^2_2}$.
The re-weighted radiation field is given by $\hat{V}_R=-\mu\hat{E}$, where $\hat{E}$ is the photon field.
Practically, the radiation field can be characterized by the so-called \emph{energy gap correlation function}
$C_R(t)={\langle}\hat{V}_R(t)\hat{V}_R(0){\rangle}$, where the time-dependence in $\hat{V}_R(t)$ denotes the interaction picture.
Moreover, we also include the decay processes from the excited state to the ground state in this study. This represents e.g.~the exciton trapping by the reaction-center in the light-harvesting complexes.

If the exciton-photon dipole coupling is sufficiently weak and the temperature of the radiation field is high, we can represent the radiation field stochastically.
Then, the dipole interaction simplifies to $\hat{H}_{S-R}=\xi\hat{K}_R$, where $\xi$ is the stochastic field.
If we define the wavefunction of the exciton as $|\psi(t){\rangle}=C_1(t)|\bar{e}_1{\rangle}+C_2(t)|\bar{e}_2{\rangle}+C_g(t)|\bar{g}{\rangle}$,
the equation of motion under the stochastic field can be shown to be
\begin{eqnarray}~\label{eom1}
\frac{dC_1(t)}{dt}&=&-i\frac{J}{\hbar}C_2(t)-(i\frac{\epsilon_1}{\hbar}+\frac{\kappa}{2})C_1(t)-i\xi_1C_g(t)\nonumber\\
\frac{dC_2(t)}{dt}&=&-i\frac{J}{\hbar}C_1(t)-(i\frac{\epsilon_2}{\hbar}+\frac{\kappa}{2})C_2(t)-i\xi_2C_g(t),\nonumber\\
\end{eqnarray}
with the radiation field $\xi_v=\frac{\mu_v}{\mu}\xi$, and the decay rate $\kappa$.
In the weak-field limit $C_g(t){\approx}1$, the Eq.~(\ref{eom1}) reduces to
\begin{eqnarray}~\label{eom2}
\frac{dC_1(t)}{dt}&=&-i\frac{J}{\hbar}C_2(t)-(i\frac{\epsilon_1}{\hbar}+\frac{\kappa}{2})C_1(t)-i\xi_1\\
\frac{dC_2(t)}{dt}&=&-i\frac{J}{\hbar}C_1(t)-(i\frac{\epsilon_2}{\hbar}+\frac{\kappa}{2})C_2(t)-i\xi_2.\nonumber
\end{eqnarray}

By using the Laplace transformation, the wavefunction coefficients and the radiation field are changed to $C_v(z)=\int^{\infty}_0e^{-zt}C_v(t)dt$ and
$\xi_v(z)=\int^{\infty}_0e^{-zt}\xi_v(t)dt$, respectively.
The equation of motion in the Laplace picture is described as
\begin{eqnarray}~\label{eom3}
\begin{pmatrix}
z+\frac{\kappa}{2}+i\frac{\epsilon_1}{\hbar} & i\frac{J}{\hbar}\\
i\frac{J}{\hbar} & z+\frac{\kappa}{2}+i\frac{\epsilon_2}{\hbar}\\
\end{pmatrix}
\begin{pmatrix}
C_1(z)\\
C_2(z)\\
\end{pmatrix}
=-i
\begin{pmatrix}
\xi_1(z)\\
\xi_2(z)\\
\end{pmatrix}.\nonumber\\
\end{eqnarray}
The Hamiltonian $\hat{H}_S$ can be diagonalized by transformation into the excitonic basis. Through the transfer matrix $\hat{S}$, it is obtained by
$\hat{S}\hat{H}_S\hat{S}^{\dag}=\sum_{v=\pm}\epsilon_{v}|\bar{e}_{v}{\rangle}{\langle}\bar{e}_{v}|$,
with $\epsilon_{\pm}$ the eigen-energy, and $|\bar{e}_{\pm}{\rangle}$ the corresponding excitonic states.
Then, the equation of motion at Eq.~(\ref{eom3}) is transformed to
\begin{eqnarray}~\label{eom4}
\begin{pmatrix}
z+\frac{\kappa}{2}+i\frac{\epsilon_+}{\hbar} & 0\\
0 & z+\frac{\kappa}{2}+i\frac{\epsilon_-}{\hbar}\\
\end{pmatrix}
\begin{pmatrix}
{C}_+(z)\\
{C}_-(z)\\
\end{pmatrix}
=-i
\begin{pmatrix}
{\xi}_+(z)\\
{\xi}_-(z)\\
\end{pmatrix}\nonumber\\
\end{eqnarray}
where the coefficients and the radiation field between the local basis and the excitonic basis are connected by $[{C}_+(z),{C}_-(z)]^T=\hat{S}[C_1(z),C_2(z)]^T$ and
$[{\xi}_+(z),{\xi}_-(z)]^T=\hat{S}[\xi_1(z),\xi_2(z)]^T$, respectively.
As a result, the expression of the time dependent coefficients at Eq.~(\ref{eom4}) in the excitonic basis are given by
\begin{eqnarray}
{C}_+(t)&=&-i\int^{t}_0e^{-(\frac{\kappa}{2}+i{\epsilon_+}/{\hbar})(t-\tau)}{\xi}_+(\tau)d{\tau},\\
{C}_-(t)&=&-i\int^{t}_0e^{-(\frac{\kappa}{2}+i{\epsilon_-}/{\hbar})(t-\tau)}{\xi}_-(\tau)d{\tau},\nonumber
\end{eqnarray}
with ${\xi}_v(\tau)=\frac{{\mu}_v}{\mu}V_R(\tau)~(v=\pm)$, which can be characterized by the correlation function as
${\langle}{\xi}_v(t){\xi}_{v^{\prime}}(0){\rangle}=\frac{{\mu}_v{\mu}_{v^{\prime}}}{\mu^2}C_R(t)$.
Hence, the population $\rho_{vv}(t)={\langle}{C}^{*}_v(t){C}_v(t){\rangle}$  at the eigen-state $|\bar{e}_{v}{\rangle}~(v=\pm)$ is expressed as
\begin{eqnarray}~\label{bpa1}
\rho_{vv}(t)&=&\int^{t}_0d{\tau_1}e^{-(\frac{\kappa}{2}-i\epsilon_v/\hbar)(t-\tau_1)}\\
&&{\times}\int^{t}_0d{\tau_2}e^{-(\frac{\kappa}{2}+i\epsilon_v/\hbar)(t-\tau_2)}{\langle}{\xi}_v(\tau_1){\xi}_v(\tau_2){\rangle}.\nonumber
\end{eqnarray}
The coherence $\rho_{+-}(t)={\langle}{C}^{*}_-(t){C}_+(t){\rangle}$ is given by
\begin{eqnarray}~\label{bpa12}
\rho_{+-}(t)&=&\int^{t}_0d{\tau_1}e^{-(\frac{\kappa}{2}-i\epsilon_-/\hbar)(t-\tau_1)}\\
&&{\times}\int^{t}_0d{\tau_2}e^{-(\frac{\kappa}{2}+i\epsilon_+/\hbar)(t-\tau_2)}{\langle}{\xi}_-(\tau_1){\xi}_+(\tau_2){\rangle}.\nonumber
\end{eqnarray}

It should be noted that this is the general solution, which does not rely on the detailed information of correlation function for the radiation field.
The only prerequisite condition is the weak-field limit. In the following, we consider the $\delta$-function noise,
which is the simplest case of the radiation field.

\subsection{$\delta$-function noise}

If the coherence time of the  radiation field is sufficitently short, the field can be considered to be Gaussian, with the corresponding correlation function specified by $C_R(t)=I^R\delta(t)$, as shown in Eq.~(\ref{cr1}). Such model is often called white noise model, or the Haken-Strobl model in the weak field limit.
Straightforwardly, the populations at states $|\bar{e}_+{\rangle}$ and $|\bar{e}_{-}{\rangle}$ are given by
\begin{eqnarray}~\label{p1}
\rho_{++}(t)=\frac{{\mu}^2_+I^R_+}{\mu^2\kappa}(1-e^{-\kappa{t}}),\\
\label{p2}
\rho_{--}(t)=\frac{{\mu}^2_-I^R_-}{\mu^2\kappa}(1-e^{-\kappa{t}}),
\end{eqnarray}
with the radiation field pumping rates $I^R_+=I^R$ and $I^R_-=I^R$.
The coherence term is given by
\begin{eqnarray}~\label{p12}
\rho_{+-}(t)=\frac{{\mu}_+{\mu}_-(I^R_++I^R_-)/2}{\mu^2(\kappa+i\epsilon/\hbar)}(1-e^{-(\kappa+i\epsilon/\hbar)t}),
\end{eqnarray}
with $\epsilon=\epsilon_{+}-\epsilon_{-}$.
From the results for the populations and the coherence, it is found that the functional expression of them will keep the same,
which implies that the physical picture is unchanged for arbitrary inter-site coupling $J$.

In absence of the decay process ($\kappa{\rightarrow}0$), the excitation is accumulated under the incoherent photon pumping.
As a result, the populations exhibit linear increase of the time.
On the other hand, the coherence  shows Rabi oscillations with finite amplitude due to the existence of the energy gap.
Hence, after a long time evolution (still in the weak-field limit), the coherence becomes negligible compared to the populations.
However, if the decay process is tuned on, both the population and coherence terms will approach the steady state, and the
static coherence occurs naturally.

\subsection{Pumping rates from quantized radiation field}

The above analysis of the dynamics for the three level molecular dimer system is based on the classical white noise,
which is also known as the Haken-Strobl-Reineker Model.
The classical radiation field can be connected with the theory with general quantum light by evaluating
the Redfield pumping rates. Specifically, we write the exciton-photon interaction Hamiltonian $\hat{H}_{S-R}$, Eqs.~(\ref{HSR-article},\ref{eq:RadiationSBC-MuE}),  in the interaction picture and the excitonic basis as
\begin{eqnarray}~\label{hsr1}
\hat{H}_{S-R}(t)&=&[\frac{{\mu}_+}{\mu}e^{i\epsilon_+t/\hbar}|\bar{e}_+{\rangle}{\langle}\bar{g}|
+\frac{{\mu}_-}{\mu}e^{i\epsilon_-t/\hbar}|\bar{e}_-{\rangle}{\langle}\bar{g}|+H.c.]\nonumber\\
&&{\otimes}\hat{V}_R(t).
\end{eqnarray}
We rewrite the radiation field quantization, Eq.~(\ref{eq:EfieldQuant}), with use of operators $V_R$ as $\hat{V}_R(t)=\sum_{k}(g_ke^{i\omega_kt}\hat{b}^{\dag}_k+g^{*}_ke^{-i\omega_kt}\hat{b}_k)$,
with $\hat{b}^{\dag}_k(\hat{b}_k)$ creation(annihilation) one photon having frequency $\omega_k$ in the momentum $k$.
The coupling constants $g_k$ are given by relation
$g_k=i\mu N_k \boldsymbol{\varepsilon}^{k}$, where   $\boldsymbol{\varepsilon}^{k}$ denotes projection of $\boldsymbol{\varepsilon}^{\mathbf{k}m}$ in the direction of molecular dipole moment, and $N_k$ is the normalization constant given at Eq.~(A4).
In this paper, the radiation field is specified as the Drude-Lorentz spectrum
$J(\omega)=\pi\sum_k|g_k|^2\delta(\omega-\omega_k)={2\hbar\lambda_R\gamma_R\omega}/{(\gamma^2_R+\omega^2)}$,
with $\lambda_R$ the coupling strength and $\gamma_R$ the cutoff frequency.
Hence, based on the second order perturbation, the pumping rate of the first excited state  is obtained by
\begin{eqnarray}~\label{Ir1}
I^R_+&=&\int^{\infty}_{-\infty}d{\tau}\sum_k|g_k|^2n_ke^{i(\omega_k-\epsilon_+/\hbar)\tau}\nonumber\\
&=&\frac{2\lambda_R\gamma_R\epsilon_+(\coth(\beta_R\epsilon_+/2)-1)}{\gamma^2_R+(\epsilon_+/\hbar)^2},
\end{eqnarray}
with the Bose-Einstein distribution $n_k=1/[\exp(\beta_R\hbar\omega_k)-1]$ and $\beta_R=1/(k_bT_R)$.
Similarly, the pumping rate for the second excited state is obtained by
\begin{eqnarray}~\label{Ir2}
I^R_-=\frac{2\lambda_R\gamma_R\epsilon_-(\coth(\beta_R\epsilon_-/2)-1)}{\gamma^2_R+(\epsilon_-/\hbar)^2}.
\end{eqnarray}
Moreover, the corresponding stimulating emission rate for the relaxation process is given by
$I^{\textrm{em}}_v={2\lambda_R\gamma_R\epsilon_v(\coth(\beta_R\epsilon_v/2)+1)}/{(\gamma^2_R+(\epsilon_v/\hbar)^2)}$.
It is found that the pumping rate and emission rate obey the detailed balance relation as $I^R_v/I^{\textrm{em}}_v=\exp(-\beta_R\epsilon_v)$.

While for the pump rate of the coherence term $\rho_{+-}$, the expression is given by
\begin{eqnarray}~\label{Ir12}
I^{R}_{+-}&=&\int^{\infty}_{0}d{\tau}\sum_k|g_k|^2n_k[e^{i(\omega_k-\epsilon_+/\hbar)\tau}+e^{-i(\omega_k-\epsilon_-/\hbar)\tau}]\nonumber\\
&=&(I^R_++I^R_-)/2.
\end{eqnarray}




\end{document}